# Controllable modification of the anisotropy energy in Laves phase YFe$_2$ by Ar$^+$ ion implantation

A. R. Buckingham[1], G. J. Bowden[1], D. Wang[1,2], G. B. G. Stenning[1], I. Nandhakumar[3], R. C. C. Ward[4] and P. A. J . de Groot[1]

1School of Physics and Astronomy, University of Southampton, Southampton, SO17 1BJ, UK.
2Department of Physics, National University of Defense Technology, Changsha, Hunan 410073, China.
3School of Chemistry, University of Southampton, Southampton, SO17 1BJ, UK .
4Clarendon Laboratory, University of Oxford, Oxford, OX1 3PU, UK.

8$^{th}$ September 2010

Implanted 3.25 keV Ar$^+$ ions have been used to modify the in-plane bulk anisotropy in thin films of epitaxially grown Laves phase YFe$_2$. The magneto optical Kerr effect, vibrating sample magnetometry and computational modeling have been used to show that the dominant source of anisotropy changes from magnetoelastic in as-grown samples to magnetocrystalline in ion implanted samples. This change occurs at a critical fluence of order $10^{17}$ Ar$^+$ ions cm$^{-2}$. The change in source of the anisotropy is attributed to a relaxation of the strain inherent in the epitaxially grown thin-films. Atomic force microscopy shows that the samples' topography remains unchanged after ion implantation. The ability to control the dominant source of magnetic anisotropy without affecting the sample surface could have important consequences in the fabrication of patterned media for high use in density magnetic data storage devices.

Due to the ever increasing demands on the density of magnetic data storage media there has been a recent influx of work on patterned media. Patterned media is expected to allow data storage densities to exceed the rapidly approaching limits of traditional granular media, thereby avoiding the superparamagnetic limit[1]. Patterned media behave as single magnetic bits, *i.e.* as single domain particles, or as a collection of strongly coupled grains, rather than as many weakly coupled grains as in existing granular media. This reduces the volume of magnetic material required per bit of information stored whilst maintaining a high signal-to-noise ratio, thereby circumventing superparamagnetism. Top-down lithographic, and bottom-up self-assembly techniques towards topographically patterned media have received much attention over the past decade[2,3]. Furthermore, a non-topographical, purely magnetic patterning process has recently become intensely investigated following its first successful demonstration in 1998[4]. Magnetic patterning, using energetic ions which either irradiate or are implanted in the media, is especially attractive since it alleviates many issues surrounding the need for planarization of the topographically patterned recording media[5,6]. In this work we report on the affects of implanting Ar$^+$ ions into epitaxially grown Laves phase YFe$_2$ samples.

The YFe$_2$ samples were grown via molecular beam epitaxy (MBE) in a *Balzers UMS 630* ultrahigh vacuum facility on epi-prepared $(11\bar{2}0)$ sapphire substrates at 600 °C. A 10 nm (110) Nb chemical buffer layer and 2 nm iron seed layer were deposited on the substrates prior to the growth of the Laves phase film[7,8]. The YFe$_2$ was grown by the co-deposition of elemental fluxes described elsewhere[9,10], following the original procedures developed by Kwo *et al.*[11]. Finally, the samples were capped with a 10 nm protective Y layer[12].

Ar$^+$ ions from an *Oxford Applied Research IG5* ion source were implanted into the YFe$_2$ samples. This was conducted under vacuum with a base pressure of $1 \times 10^{-6}$ mBar, which rose to $1 \times 10^{-4}$ mBar during the implantation process. The samples were continuously rotated with the incident Ar$^+$ ions making an angle of 68° to the sample surface normal. The fluence of Ar$^+$ ions was calculated from measurements of the current density induced at the sample surface (1 $\mu$Acm$^{-2}$) and the acceleration voltage (3.25 keV). The incident fluence ranged from zero through to $1.7 \times 10^{17}$ Ar$^+$ ions cm$^{-2}$. Sample characterization was performed by longitudinal magneto optical Kerr effect (MOKE) measurements, vibrating sample magnetometry (VSM) (*Oxford Instruments Aersonic 3001*) and atomic force microscopy (AFM) (*Digital Instruments Multimode SPM*) both before and after implantation with various Ar$^+$ ion fluences. Modelling of the implantation of ions was conducted using SRIM[13] whilst modelling of the sample magnetization was performed with OOMMF[14].

MOKE magnetization data for a sample in an as-grown state is presented in graphs a) and b) in Figure 1. These data clearly show that the sample exhibits an anisotropic response to an applied magnetic field. From the hysteresis loop obtained with the magnetic field applied along the $[\bar{1}10]$ direction (graph a)), the sample has a coercivity $H_C$ = 17.4 mT and a squareness $\frac{M_R}{M_S}$ = 1.0 (with $M_R$ and $M_S$ the magnetization at remanence and saturation, respectively). In graph b) the hysteresis loop obtained with the magnetic field applied along the [001] direction is presented; here both the coercivity and squareness are greatly reduced, to $H_C$ = 5.2 mT and $\frac{M_R}{M_S}$ = 0.2 respectively. This corresponds to a hard axis of magnetization where the cubic anisotropy of the Laves phase structure is clear. There is good qualitative and quantitative agreement between these data and the VSM data for the same sample (Figure 2, graphs a) and b)), confirming that the observed anisotropy is a bulk, rather than surface effect. Note that the skin depth of YFe$_2$ ($\delta_{\text{YFe}_2}$) at 633 nm (HeNe laser used for MOKE measurements) is approximately 20 nm using values for electrical resistivity at room temperature from Ikeda and Nakamichi[15]).

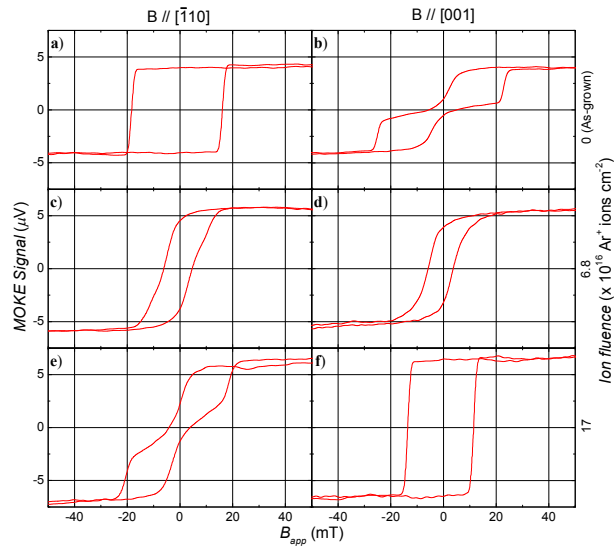

Figure 1: (Colour online) MOKE hysteresis loops of a 100 nm thick $YFe_2$ sample in the as-grown state (graphs a) and b)) and for samples implanted with increasing fluences of $Ar^+$ ions (graphs c) – f)). The magnetic field ($B_{app}$) has been applied along two orthogonal directions as indicated by the upper labels.

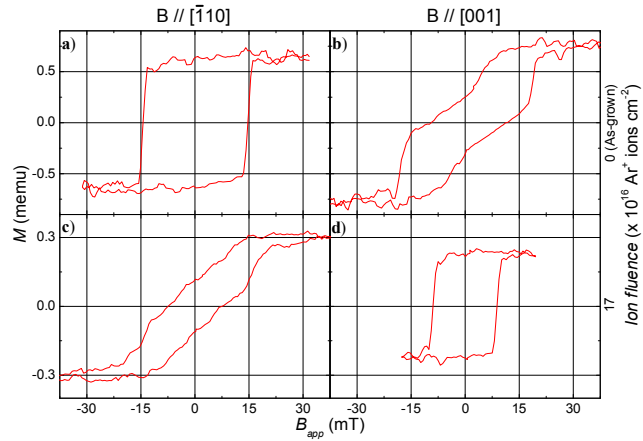

Figure 2: (Colour online) VSM hysteresis loops of a 100 nm thick $YFe_2$ sample in the as-grown state (graphs a) and b)) and for a sample implanted with a fluence of $1.7 \times 10^{17}$ $Ar^+$ ions $cm^{-2}$ (graphs c) and d)). The magnetic field ($B_{app}$) has been applied along two orthogonal directions as indicated by the upper labels.

It has been reported that epitaxially grown Laves phase $YFe_2$ exhibits little or no magnetoelastic anisotropy[29,16] since it is usually described within the context of multilayer exchange-spring $YFe_2$/$REFe_2$ magnetic materials (*e.g.* RE = Dy, Er)[17,18,19,20], which exhibit a considerably larger anisotropy than the soft $YFe_2$ alone. In these $REFe_2$ materials the anisotropy is ascribed to second order crystal field terms originating from the RE atoms in the Laves phase structure. Previously Wang *et al.*[21] presented anisotropic magnetization data for strained Laves phase $YFe_2$ samples, noting its evolution due to the increased dominance of magnetostatic effects induced by micron-scale patterning. Combining these data with the hysteresis loops in Figure 1 and Figure 2 provides conclusive evidence that there is a significant anisotropy energy associated with the Laves phase YFe$_2$.

It is well established that during the first stages of deposition in epitaxial REFe$_2$ systems the growth mechanism is island-like[10,22], becoming continuous once the REFe$_2$ thickness reaches 25 nm. Island growth is a classical signature of a growth mechanism which is evolving in an attempt to reduce strain within the system[23,24]. Indeed, the lattice parameters of bulk Laves phase REFe$_2$ materials differ significantly from those of the MBE grown samples. X-Ray diffraction data has confirmed that there is an in-plane expansion and perpendicular contraction in the Laves phase unit cell[16]. Accordingly the MBE grown samples are said to be strained and thus are subject to a magnetoelastic energy term, the origin of which is thermal in nature, induced during the post-growth cooling of the sample from 600°C to room temperature[16]. In the bulk, un-strained Laves phase REFe$_2$ compounds there is a dominant magnetocrystalline anisotropy energy ($E_{MC}$)[25,26,27], defining the easy and hard axes of magnetization. However, in MBE grown thin-films of Laves phase REFe$_2$ compounds, the magnetoelastic energy ($E_{ME}$) induced during the sample growth is larger than $E_{MC}$, serving to alter the easy and hard axes of magnetization[22]. It has been shown that the MBE grown Laves phase samples are subject to a shear strain of $\varepsilon_{xy} = -0.5\%$[16]. This is determined experimentally to be the dominant strain term, thus the standard equation describing $E_{ME}$ in a system with cubic anisotropy (*e.g.* see refs. 29 and 28) may be given by a much simpler relationship;

$$E_{ME} = b_2 \varepsilon_{xy} \alpha_x \alpha_y \quad (1)$$

where $b_2$ is a magnetoelastic coefficient given in ref. 29 and $\alpha_x$ and $\alpha_y$ are direction cosines. In thin-film samples of YFe$_2$, $E_{ME}$ is expected to dominate. The easy axis of magnetization is shown to be along equivalent ⟨221⟩ directions[29], close to the [$\bar{1}$10] direction shown in graphs a) of Figure 1 and Figure 2.

After implantation by a fluence of > $10^{17}$ Ar$^+$ ions cm$^{-2}$ (graphs e) & f) in Figure 1 and graphs c) and d) in Figure 2), it is clear that the easy and hard axes of magnetization have exchanged directions from those in the as-grown samples. From the MOKE data for the sample implanted with a fluence of $1.7 \times 10^{17}$ Ar$^+$ ions cm$^{-2}$, the coercivity is reduced to $H_C$ = 4.0 mT and the squareness to $\frac{M_R}{M_S}$ = 0.3 when the magnetic field is applied along the [$\bar{1}$10] direction. When the field is applied along the [001] direction, the coercivity is increased to $H_C$ = 12.5 mT and squareness to $\frac{M_R}{M_S}$ = 1.0. The decreases along the [$\bar{1}$10] direction and the increases along the [001] direction, and the accompanying change in shape of the hysteresis loops, corresponds to a rotation of 90° in the easy and hard axis of magnetization within the ion implanted sample. After ion implantation above a critical fluence of approximately $10^{17}$ $Ar^+$ ions $cm^{-2}$, there is a close to easy axis of magnetization along the [001] direction, whilst the hard axis lies close the [$\bar{1}$10] direction. It is worthwhile noting that at a fluence of 6.8 × $10^{16}$ $Ar^+$ ions $cm^{-2}$ (graphs c) and d), Figure 1), the hysteresis loops obtained for the magnetic field applied along orthogonal directions are equivalent. This gradual progression to the rotation of the easy and hard axes of magnetization introduces a useful degree of controllability into engineering the magnetic properties via ion implantation.

Furthermore, we note that the VSM data for the ion implanted sample agrees with the MOKE data. From this we may conclude that the apparent re-orientation of the magnetic axes is throughout the sample, rather than being confined to a small region at the top of the sample corresponding to the MOKE skin depth, although some material is removed via ion milling as illustrated by the reduction in $M_S$ in the VSM data. These data show a reduction in $M_S$ of ~60% (0.65 to 0.25 memu for the easy axes of magnetization (graphs a) and d)) and 0.80 to 0.30 memu for the hard axes of magnetization (graphs b) and c))), corresponding to a reduction in sample thickness to ~40 nm. This is consistent with the MOKE data, i.e. $\delta_{YFe_2}$ < 40 nm. Ion milling also accounts for the reduced amplitude of the MOKE signal in Figure 1, graphs a) and b). The protective Y cap is removed via ion milling, thus the ion implanted samples exhibit a larger amplitude of MOKE signal due to the experiment probing a larger volume of the magnetic material. Note that the MOKE signal amplitude does not begin to decrease, showing that the sample thickness remains greater than $\delta_{YFe_2}$.

OOMMF modelling of 100 nm thick layers of strained $YFe_2$ has been conducted. A one dimensional spin-chain model was utilised with a mesh size of 1 $nm^3$. The model was tailored to include a strain pre-factor ($S_{PF}$)[20] in order for the theoretical work of Bowden et al.[30] – which appears to under estimate the manetoelastic strain – to agree with the experimental work of Zhukov et al.[31]. Hysteresis loops from OOMMF calculations to model an as-grown sample are presented in Figure 3, graphs a) and b), with the [$\bar{1}$10] and [001] directions corresponding to the easy and hard axes respectively. OOMMF simulations were also performed with reduced $S_{PF}$ values in order to model the effects of a reduced $E_{ME}$. These hysteresis loops, generated with $S_{PF}$ = 1.0 and 0.0, are shown in Figure 3, graphs c) – f). The data in graphs e) and f) are the direct opposite of those in graphs a) and b); the easy and hard axes of magnetization are now along the [001] and [$\bar{1}$10] directions respectively. This is in good agreement with the experimental MOKE and VSM data. Additionally, an intermediate step is also apparent in Figure 3 (graphs c) and d)), where the OOMMF calculated hysteresis loops show that the response of the sample's magnetization to $B_{app}$ to be almost equivalent whether it is applied along the [$\bar{1}$10] or [001] directions, just as evidenced in the MOKE data for a sample subject to a fluence of $6.8 \times 10^{16}$ Ar$^+$ ions cm$^{-2}$. As the strain within the sample is reduced, $E_{ME}$ becomes less dominant until eventually $E_{MC}$ dominates, with the two being approximately equal at some intermediate crossing point.

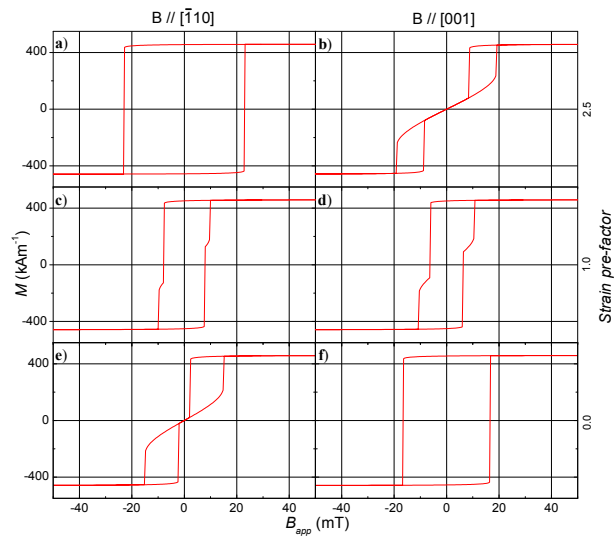

Figure 3: (Colour online) OOMMF calculated data for 100 nm thick YFe$_2$ layer subject to a variation in the dominance of the magnetoelastic anisotropy energy. Graphs a) and b) represent a YFe$_2$ sample in its as-grown, strained state. Graphs c) – f) show the corresponding hysteresis loops as the magnetoelastic anisotropy energy is reduced. The magnetic field ($B_{app}$) has been applied along two orthogonal directions as indicated by the upper labels.

AFM was used to characterize the sample topography both in the as-grown state and after implantation by a fluence of $6.8 \times 10^{16}$ Ar$^+$ ions cm$^{-2}$. The AFM micrographs are presented in Figure 4, images a) and b). Although a certain amount of material is inevitably sputtered away during the ion implantation process, from these data it is clear that the implanted Ar$^+$ ions have had no other effects on the sample topography. The surface roughness remains as $3 \pm 0.3$ nm. From this we may conclude that the reorientation of easy and hard axes of magnetization are not due to an ion implantation induced shape anisotropy.

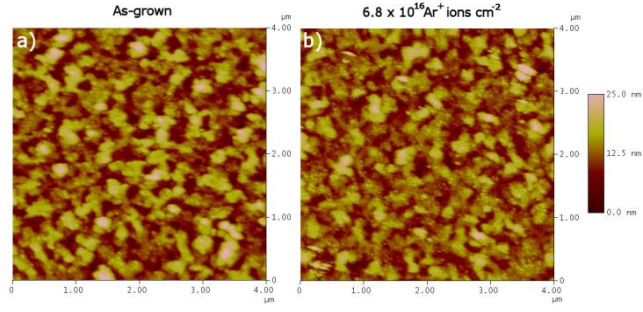

Figure 4: (Colour online) AFM micrographs of a YFe$_2$ sample in the as-grown state (image a)) and after implantation by a fluence of $6.8 \times 10^{16}$ Ar$^+$ ions cm$^{-2}$ (image b)).

Implantation profiles for 3.25 keV Ar$^+$ ions have been calculated using `SRIM`. The results confirm that all of the Ar$^+$ ions are stopped within 10 nm of the sample surface. Thus despite the effects of ion milling, a volume significantly larger than that penetrated by the Ar$^+$ ions remains after the implantation process, yet the effects of these Ar$^+$ ions are felt throughout the sample. The Ar$^+$ ions energy is deposited in the YFe$_2$ sample via predominantly elastic collisions; `SRIM` calculations show that the largest proportion of which is to phonons created within the crystalline lattice. The incident Ar$^+$ ions suffer many recoil collisions with the YFe$_2$ lattice, but recoil cascades are rare due to the large mass difference between the incident ion and the tightly bound target atom[32]. Therefore the overall crystalline structure of the Laves phase is maintained, with the recoiled YFe$_2$ atoms relaxing only to their un-strained lattice positions. From the anisotropic magnetization data it is clear that a well ordered crystalline structure remains throughout the material after Ar$^+$ ion implantation. We suggest that it is due to the regular epitaxial structure of the Laves phase sample, that the effects of ion implantation are much longer-range than predicted by `SRIM` modelling. The recoils induced within the YFe$_2$ crystalline lattice, which incidentally is a relatively brittle material, propagate readily throughout the entire epitaxial structure via the rapid propagation of ion implantation-induced dislocations, in much the same way a crack readily propagates through glass.

In summary we present the first demonstration of ion implantation induced changes to the magnetic properties of epitaxial Laves phase YFe$_2$. Previous work has demonstrated the ability to induce a uniaxial anisotropy in polycrystalline films via ion implantation[33,34], and also substantial work exists on modification of the plane in which the easy axis of magnetization lies by ion irradiation induced modifications to interfacial anisotropy (*e.g.* see refs, 4, 35 and 36). However, this work is the first to demonstrate ion implantation as a tool to engineer the dominant anisotropy energy of an epitaxial Laves phase system. Moreover we show that the change from a magnetoelastic to a magnetocrystalline anisotropy dominated system can be accurately controlled by simply varying the incident ion fluence. Such a controllable way of tailoring magnetic properties could be useful for the fabrication of patterned magnetic media when combined with the use of lithographically defined stencils, eliminating both media planarization processes[5,6] and the proximity effect[37].